\documentstyle[psfig]{ioplppt}

\newcommand{\R}{{\bf R}}
\newcommand{\cA}{{\cal A}}
\newcommand{\cW}{{\cal W}}
\newcommand{\cM}{{\cal M}}
\newcommand{\supp}{\mbox{supp}}
\newcommand{\matr}[4]{\left(\begin{array}{cc}{#1}&{#2}\\{#3}&{#4}\end{array}\right)}
\newcommand{\vect}[2]{\left(\begin{array}{c}{#1}\\{#2}\end{array}\right)}

\newtheorem{theorem}{Theorem}
\newtheorem{corollary}{Corollary}
\newtheorem{lemma}{Lemma}

\newcommand{\proof}{\noindent{\bf Proof: }}
\newcommand{\eproof}{\noindent QED \vspace{3mm}}

\begin{document}
\jl{8}

\title{Transverse instability for non-normal parameters}
\author{Peter Ashwin\dag\ftnote{3}{E-mail:P.Ashwin@mcs.surrey.ac.uk}, 
Eurico Covas\ddag\ftnote{4}{E-mail:eoc@maths.qmw.ac.uk} 
and Reza Tavakol\ddag\ftnote{5}{E-mail:reza@maths.qmw.ac.uk}}

\address{\dag\  Department of Mathematical and Computing Sciences,
                University of Surrey,
                Guildford GU2 5XH, UK}
\address{\ddag\ Astronomy Unit,
                School of Mathematical Sciences,
                Queen Mary \& Westfield College,
                Mile End Road,
                London E1 4NS, UK}
\maketitle
\begin{abstract}
Suppose a smooth dynamical system has an invariant subspace and
a parameter that leaves the dynamics in the invariant subspace
invariant while changing the normal dynamics. Then we say the parameter
is a normal parameter, and much is understood of how attractors can
change with normal parameters. Unfortunately, normal parameters do
not arise very often in practise.

We consider the behaviour of attractors near invariant subspaces on
varying a parameter that does not preserve the dynamics in
the invariant subspace but is otherwise
generic, in a smooth dynamical system. We refer to such a parameter as
``non-normal''.  If there is chaos in the invariant subspace that is
not structurally stable, this has the
effect of ``blurring out'' blowout bifurcations over a range of
parameter values that we show can have positive measure in parameter space.

Associated with such blowout bifurcations are bifurcations to
attractors displaying a new type of intermittency that is
phenomenologically similar to on-off intermittency, but where the
intersection of the attractor by the invariant subspace is larger than
a minimal attractor. The presence of distinct repelling and
attracting invariant sets leads us to refer to this as
``in-out'' intermittency. Such behaviour cannot appear
in systems where the transverse dynamics is a skew product over the
system on the invariant subspace.

We characterise in-out intermittency
in terms of its structure in phase space and in terms of invariants of
the dynamics obtained from a Markov model of the attractor.  This model
predicts a scaling of the length of laminar phases that is similar to
that for on-off intermittency but which has some differences.

Finally, we discuss some other bifurcation effects associated with
non-normal parameters, in particular a bifurcation to riddled basins.
\end{abstract}

\section{Introduction}
Nonlinear dynamical systems with invariant subspaces forced by, for
example, symmetries or other constraints are of great physical
interest. The dynamics on such invariant subspaces can show transverse
stability or instability depending on whether small perturbations away
from the subspace decay or grow with time. As these subspaces need not
be normally hyperbolic, the dynamics near them can be very complicated;
for example, the phenomena of riddled
basins is typical in such systems \cite{Ale&al92}.

Recently there have been concerted attempts to understand the
bifurcations of such attractors on varying a parameter in the system.
For system parameters that are {\em normal} \cite{AshBueSte96} i.e.
parameters that leave the system on the invariant manifold unchanged,
there is a good description of the instability that causes a blowout
bifurcation \cite{Ott&Som94}; this is a linear instability
that can be computed by examining the Lyapunov exponents (LE) corresponding
to perturbations in transverse directions. The problem of characterising the
global branching in such bifurcations is still not understood, but some
progress has been made in \cite{AshAstNic97}.

In reality, most physically relevant parameters are not
normal; that is, they vary the dynamics within the invariant subspace
as well as that outside it. This, coupled with the fact that most physically
relevant chaos is not structurally stable (it is {\em fragile} in the
terminology of \cite{Bar&al97,Tav&Ell88}), and the fact that most
systems do not possess a skew product nature, leads to a variety of
novel and physically relevant phenomena that we investigate in
this paper. These include

\begin{enumerate}
\item ``Blurring'' of a blowout bifurcation caused by the breakup of fragile
chaos.
\item Windows within parameter space where a generalised type of on-off
intermittency (which we call ``in-out'' intermittency) may appear.
\item New mechanisms for bifurcation to locally riddled basins,
including cases where the attractor in the invariant subspace is not
chaotic.
\end{enumerate}
All of these phenomena arise on varying parameters that are not
normal in systems that are fragile and do not have skew product form.

More precisely, consider a family of smooth mappings
$$
f_\nu:M \rightarrow M
$$
parametrised smoothly by $\nu\in\R$,
where $M$ is a compact subset of $\R^m$ with open interior,
such that $f_\nu$ leaves a linear
subspace $N\subset M$ invariant. If the map $h=f_{\nu}|_N$,
restricted to $N$, is independent of $\nu$ we say that $\nu$ is a {\em
normal parameter} for $f$ on $N$, otherwise we say $\nu$ is a {\em
non-normal parameter}.

If a (minimal Milnor) attractor $A\subset N$ becomes
transversely unstable on varying a parameter $\nu$
we call this a {\em blowout bifurcation}. For normal parameters
this can happen at isolated values of $\nu$;
we will show that for non-normal parameters a more typical scenario is the
{\em blurred blowout bifurcation}, where blowouts can accumulate on
themselves and can even occur on a positive measure subset of
parameters. In practice, a blurred blowout is recognisable as a
complicated pattern of inflating and collapsing of the basins of attraction
of a family of attractors within $N$.

Now suppose that $A$ is a chaotic attractor for $f$ at some particular
parameter value, and $A\not\subset N$. Suppose moreover that $A_0=A\cap N$ is
non-empty and so there are trajectories in $A$ that get arbitrarily
close to $N$ but also a finite distance from $N$, arbitrarily many
times. In the case that $A_0$ is a (minimal) attractor for $f|_N$, we
say that $A$ displays on-off intermittency to the attractor $A_0$ in $N$
\cite{Pla&al93}. The attractor $A$ is {\em stuck on} \cite{Ash95} to
$A_0$ for $f|_N$.

We will see that in general, although $A_0$ is an attractor,
it need not be a minimal attractor
for $f|_N$; in this case we have ``transversely attracting'' and
``transversely repelling'' invariant subsets within $A_0$; this more
general case we refer to as {\em in-out} intermittency.  If we
have an in-out intermittent state that is not on-off, then a minimal
attractor for $f|_N$ is a proper subset of $A_0$; these
can be, for example, stable periodic orbits within $N$.

This form of intermittency manifests itself as an attractor where
trajectories show long periods close to $N$ shadowing orbits in
$A_0$, alternating with short bursting phases that may or may not be
transient. In particular the ``growing'' and the ``decaying'' phases
can happen via different mechanisms within the invariant subspace;
in particular, {\em only the phases where the trajectory moves
away from $N$ remain close to an attractor within $N$}.

This sort of intermittency is a truly global phenomenon in that it
{\em cannot} appear in systems with skew-product structure, even if
they show blowout bifurcations etc. Since most systems with invariant
subspaces do not have skew-product structure, we therefore believe that
in-out intermittency will be commonly observable.

We discuss in Section~\ref{secmodel} a simple model mapping of the
plane with a non-normal parameter. This mapping can be analysed
fairly comprehensively, both numerically and theoretically. It can
be shown to demonstrate all of the above (and many more) phenomena.
In Section~\ref{sectheory} we put forward arguments to show that the
observed behaviour of the map in Section \ref{secmodel} is in fact
typical. Essential to our arguments are the use of Lyapunov
exponents (LEs), minimal Milnor attractors, fragility of
chaotic attractors and the lack of a skew product structure.

Section~\ref{secinout} discusses and characterises in-out intermittency
in more detail, and shows that the essential features of
this behaviour are well-modelled by a Markov map. This map
has two parameters that can be used to characterise the
intermittency. We define a ratio $R_{io}$ of
time spent in the ``in'' and ``out'' phases. In the limit of
$R_{io}=0$ we regain on-off intermittency. This quantity is
an invariant of the intermittent dynamics up to conjugacy.

Next, Section~\ref{secbifs} examines other bifurcation effects that
appear in systems on varying non-normal parameters; notably we
find a new route to creation of a riddled basin via a ``non-normal''
riddling bifurcation that may well be more common than that
described in \cite{Ott&al94}.

Finally in Section \ref{secdiscuss} we discuss our results and
point out similar behaviour that has previously been found in
both ordinary and partial differential equation models.

\section{A model planar mapping with a non-normal parameter}
\label{secmodel}

\subsection{The model}

The model we consider extends the well-known logistic map
to a mapping of the plane
\begin{equation}\label{eqmap}
f(x,y)=( rx(1-x)+s xy^2, \nu e^{bx}y+ay^3).
\end{equation}
This has five parameters $r\in[0,4]$ and $(s,\nu,a,b)\in\R^4$ and it displays
a wide variety of bifurcation behaviour. We can view this as
a map of $\R^2$ to itself that leaves
$N=\R\times\{0\}$ invariant (in fact for arbitrary parameter values
most orbits diverge to infinity. Nevertheless, for $0<r<4$ there is a
compact subset of initial conditions containing $N$ that remains
bounded). The map on $N$ is the well-known logistic map and
it undergoes the familiar routes to chaos via intermittency and
period-doubling cascades.

In the case $s=0$, the map has extra structure; it is a {\em skew
product} over the dynamics in $x$, i.e.\ it can be written
as
\begin{equation}\label{eqskewprod}
f(x,y)=(h(x),g(x,y))
\end{equation}
for some $g$ and $h$, where $x\in N$. We will see that the breaking
of the skew product form ($s\neq 0$) is important to see the generic types
of dynamics we report here.

If we fix $r$ and vary $s$, $\nu$, $a$ and $b$ we see that the latter
four parameters do not affect the map restricted to $N$; i.e.\ $s$,
$\nu$, $a$ and $b$ are {\em normal parameters} for the system
restricted to $N$. We are especially interested in the case
where these parameters are fixed and the only {\em non-normal
parameter} $r$ is varied.
In this case the dynamics in $N$ will undergo many bifurcations in
regions of interest.

We investigate the relationship between the dynamics and the
numerically measured LEs; these are
defined as usual for $(x,y)\in\R^2$ and $(0,0)\neq(u,v)\in\R^2$ by
$$
\lambda_{(x,y)}(u,v)=\lim_{n\rightarrow\infty} \frac{1}{n}
\|\log Df^n_{(x,y)}(u,v)\|
$$
where this limit exists \cite{Ose68}; in Section~\ref{sectheory} we
will return to explain the dynamical behaviour in terms of these
rates of asymptotic separation of trajectories.

Roughly speaking, $\lambda_{(x,y)}(u,v)$ measures the exponential rate of
growth of perturbations in the direction $(u,v)$ along the orbit
of $(x,y)$; typically there will be two distinct Lyapunov exponents.

If $(x,y)\in N$, i.e.\ $y=0$ then we can classify the LE
into two cases: the {\em tangential} LE
$$
\lambda_{||}(x)=\lambda_{(x,0)}(u,0)
$$
and the {\em transverse } LE
$$
\lambda_{\perp}(x)=\lambda_{(x,0)}(u,v)
$$
for some $(u,v)$ such that $v\neq 0$. (Note that it may be the
case that $\lambda_{(x,0)}(u,v)=\lambda_{||}(x)$ for almost all
$(u,v)$; however if it is different for one value of $(u,v)$
this is $\lambda_{\perp}$).

\subsection{Numerical experiments}
To get a qualitative understanding of the dynamics possible
in non--normal, fragile and skew product settings, we numerically calculated a
number of dynamical indicators for the system (\ref{eqmap})
over a range of control parameters in
order to investigate the corresponding attractors and their
relationship to the invariant subspace $N$. We now summarise some of the
most interesting types of new behaviour we have observed.

\subsubsection{Non-normal blowout bifurcation}

\begin{figure}
\begin{center}\mbox{\psfig{file=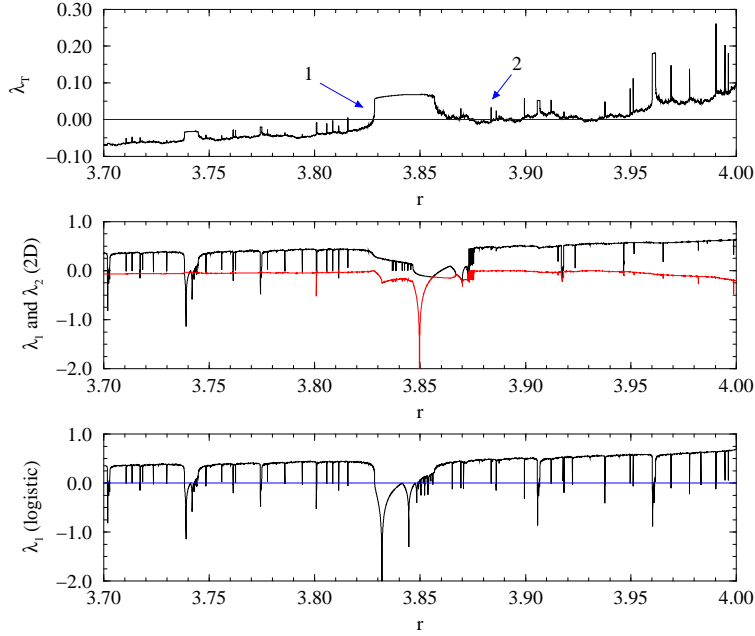,width=10cm}}\end{center}
\caption{\label{figliapscan}
Numerically computed LEs of attractors for
the system (\protect{\ref{eqmap}}), upon changing
the non-normal parameter $r$. Here
$\lambda_T$ and $\lambda_1$ (logistic) are respectively the transverse and tangential
exponents for an attractor in $f|_{N}$ (i.e.\ with initial
condition $y=0$), and $\lambda_{1}$ and $\lambda_2$ (2D) refer to
the LEs for an attractor of the full system.
The regions marked by 1 and 2 in the top panel correspond to
different occurrences of ``blurred blowout bifurcations'', where $\lambda_T$ passes
through $0$. The other parameters are fixed at $(s,\nu,a,b)=(-0.3,1.82,-1,-1)$}
\end{figure}

Figure \ref{figliapscan} shows the numerically computed
LEs, for the full system (\ref{eqmap}) as well as the LE
corresponding to the map $h(x)=f|_N$ restricted to $N$ and the
transverse LE, $\lambda_T$, as a function of the parameter $r$.  In
the top panel we have marked by 1 and 2 two different regions
where there are ``blowout'' bifurcations.
We show in Figure \ref{figatts1}
typical attractors of system (\ref{eqmap}) with
parameter values in these regions for some
initial conditions.

\begin{figure}
\begin{center}

(a)

\hspace{5mm}


(b)

\end{center}
\caption{\label{figatts1}
Panel (a) represents the phase space
 picture for the full system (\protect{\ref{eqmap}}) for parameter
values $r=3.8285$, $\nu=1.82$, $s=-0.3$, $a=-1$ and $b=-1$ (in the
region marked 1), and initial conditions $x=1/2$, $y=1/2$,
whereas panel (b) represents a similar picture
(corresponding to
region marked 2) with
parameter
values $r=3.88615$, $\nu=1.82$, $s=-0.3$, $a=-1$ and $b=-1$
and initial conditions $x=1/2$, $y=1/2$. }
\end{figure}

We have also calculated in each case the average distances and maximum
distance of typical trajectories from $N$ and LEs
after transients have been allowed
to die away. These are shown in
Figures~\ref{figscan} and \ref{figscanblur}.

The regime 1 corresponding to
$r=3.828$ is brought about by a saddle-node going to a stable period
$3$ orbit in $N$ that breaks down to chaos on reducing $r$ via
saddle-node induced intermittency.
The regime 2 located near $r=3.886$ seems to have attractors that
barely vary their
statistics.

In both cases the blowout is observable over a small range of $r$.
As shown in Figures~\ref{figscan}(d) and \ref{figscanblur}(d),
the transverse LE appears to vary non-differentially over a large
measure set of $r$. We suppose that these show essentially
the same phenomenon, with just a different
measure set of points where `blowout' bifurcations occur.

\begin{figure}
\caption{\label{figscan}
Details of the LEs (a,c,d) and the average and maximum
distances (b) of a typical trajectory from the invariant manifold $N$ for
attractors of the map ({\protect \ref{eqmap}}), in the region marked
1. The minimum distance is always very close to zero and hence has
been excluded. (d) shows a seemingly non--differentiable change in transverse
Lyapunov exponent as a function of $r$, organised by a saddle-node to
intermittency at $r=3.828$. Note however the ``spikes'' in $\lambda_T$
in the chaotic regime, $r<3.282$, indicating the locking onto large
periodic orbits that are very close in invariant measure to the period
three orbit existing for $r>3.282$.
Increasing the number of iterates
does not seem to affect the structure of this figure.
The other parameters are fixed at $(s,\nu,a,b)=(-0.3,1.82,-1,-1)$.}
\end{figure}

\begin{figure}
\caption{\label{figscanblur}
Details of the LEs (a,c,d) and the average and maximum
distances (b) of a typical trajectory from the invariant manifold $N$ for
attractors of the map ({\protect \ref{eqmap}}), in the region marked
2, where the transients have been allowed to decay.
The minimum distance is always very close to zero and hence has been
excluded, implying that all attractors are ``stuck on'' to $N$ if they are
not actually within $N$.  (d) shows an apparently
non--differentiable change in transverse Lyapunov exponent as a
function of $r$, resulting in a blurred blowout near
$r=3.8866$. Increasing the number of iterates
does not seem to affect the structure of this figure.
The other parameters are fixed at $(s,\nu,a,b)=(-0.3,1.82,-1,-1)$.}
\end{figure}

\begin{figure}
\caption{\label{figscan2dim}
Successive scan with amplifications through the two dimensional parameter space $(\nu,r )$, showing
regions of transverse stability (below the boundary curve) and transverse
instability (above the boundary curve) of typical attractors in $N$.
The other parameters are fixed at $(s,a,b)=(-0.3,-1,-1)$. This boundary has
locally the form of a graph over $r$ because $\nu$ is a normal parameter.}
\end{figure}

Finally, Figure \ref{figscan2dim} shows the boundaries of the regions in the
2--dimensional $(r,\nu)$ parameter space for which the attractor in
$N$ is transversely stable (unstable).  Note how there are two regions
in this parameter space; one in which there is an attractor contained in
the invariant subspace and another where there is not. The boundary
between the two takes the form of a graph over $r$. This is because
$\nu$ is a normal parameter and so for a given attractor in $N$ this
varies the normal LE smoothly through zero.  The lack of normality
of $r$ means that the variation is much more complicated in this
direction.

These numerical computations demonstrate a number of important features
of the blowout bifurcations observed here, notably:

\begin{itemize}
\item[(a)] There are oscillations through zero of the transverse
Lyapunov exponent of the attractor for the map $h$.
\item[(b)] The set of parameter values where the tangential LE
is positive (corresponding to a chaotic attractor for $h$) are
interrupted by a large number of intervals where the LE
is negative and the dynamics is periodic.
\item[(c)] The minimum distance from $N$ is
always observed to be exactly (or very close to) zero,
implying that all attractors are ``stuck on'' to $N$ if they are
not actually within $N$.
\end{itemize}

On the basis of results in the next section, we conjecture that there is
in fact a positive measure set of $r$ that correspond to
blowout bifurcations in some sense.

\subsubsection{Intermittent behaviour}

\begin{figure}
\begin{center}
\mbox{\psfig{file=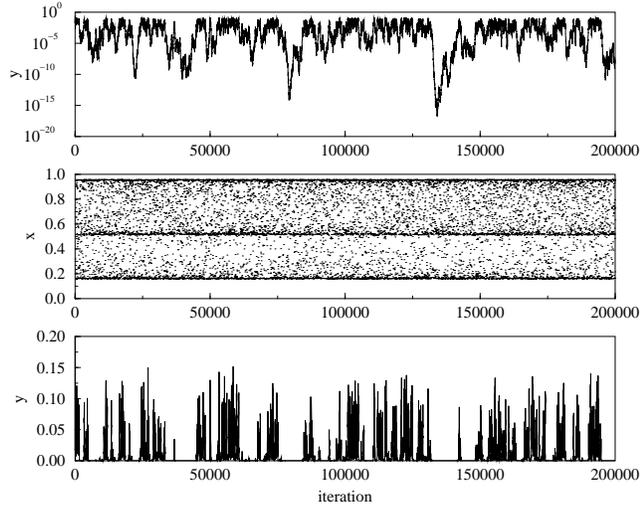,width=85mm}}

(a)

\hspace{5mm}

\mbox{\psfig{file=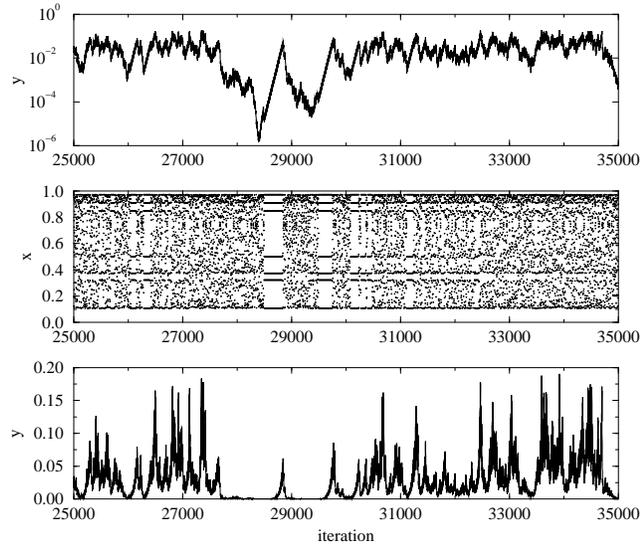,width=85mm}}

(b)
\end{center}
\caption{\label{figinoutonoff}
Panels (a) and (b) show examples
of time series generated by iterating (\protect{\ref{eqmap}})
from a randomly chosen initial condition, after transients have
been allowed to die out. In each case, the top plot shows $y$ in a logarithmic scale,
and the second and the third plots show
$x$ and $y$ on linear scales.
Panel (a) shows an example of on-off intermittency at $r=3.82786$
($\lambda_T=0.0024$) and panel (b) an example of what we refer to as
in-out intermittency at $r=3.88615$ ($\lambda_T=0.023$).
The other parameters are fixed at $(s,\nu,a,b)=(-0.3,1.82,-1,-1)$.
}
\end{figure}

Figure~\ref{figinoutonoff} shows two examples of intermittent-type
behaviour for the system ({\protect \ref{eqmap}}), involving an
attractor that is not in $N$ but which contains points arbitrary close
to $N$. Figure ~\ref{figinoutonoff}(a) shows an example of the well-known on-off
intermittency \cite{Pla&al93}; while Figure ~\ref{figinoutonoff}(b) shows an example of a
related intermittency (in-out intermittency) where the attractor in $N$ is much smaller than
the attractor for the full system intersected with $N$.
In the latter case, the
statistics of the attractor on $N$ are markedly different from the
statistics of the attractor for the full system near $N$. This is reflected by
the ``windows''  of periodic behaviour in the $x$ dynamics when $y$ is small
and growing exponentially over several orders of magnitude.
Section \ref{secinout} presents a theoretical analysis and characterisation
of this type of intermittent behaviour.

\subsubsection{Non-normal riddling bifurcation}

Suppose there is a minimal attractor $A\subset N$ with a (locally) riddled basin
\cite{Ale&al92,AshBueSte94} that is fragile. On varying a non-normal parameter
it is possible for this to collapse onto a stable periodic orbit
within $N$ under arbitrarily small perturbations. If the periodic
orbit is linearly stable we can infer that its basin of attraction has open
interior and so upon variation of a non-normal parameter
we can get transitions from open basins
to and from riddled basins in a very natural way.

\begin{figure}
\begin{center}
\mbox{\psfig{file=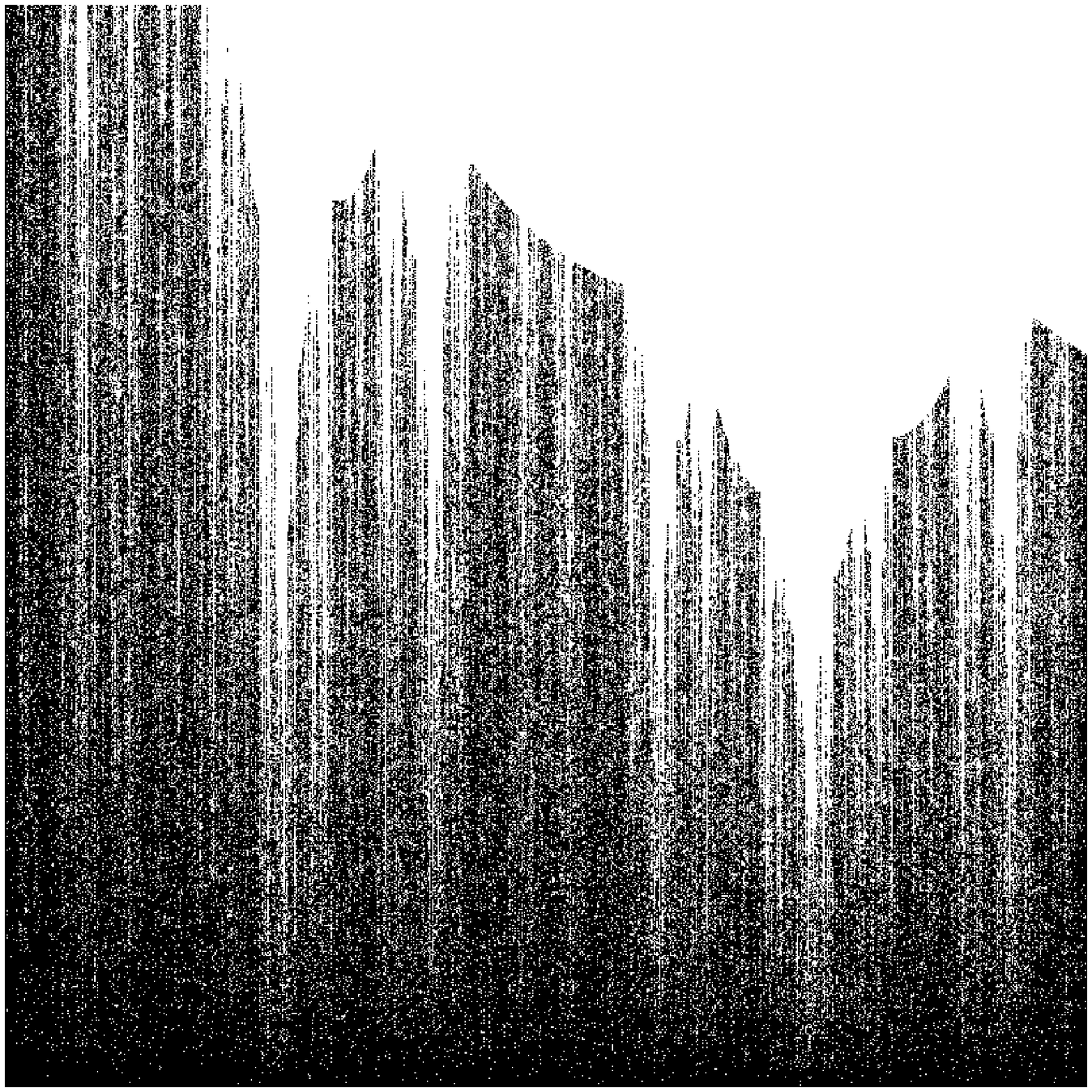,width=6cm}}

(a)

\hspace{5mm}

\mbox{\psfig{file=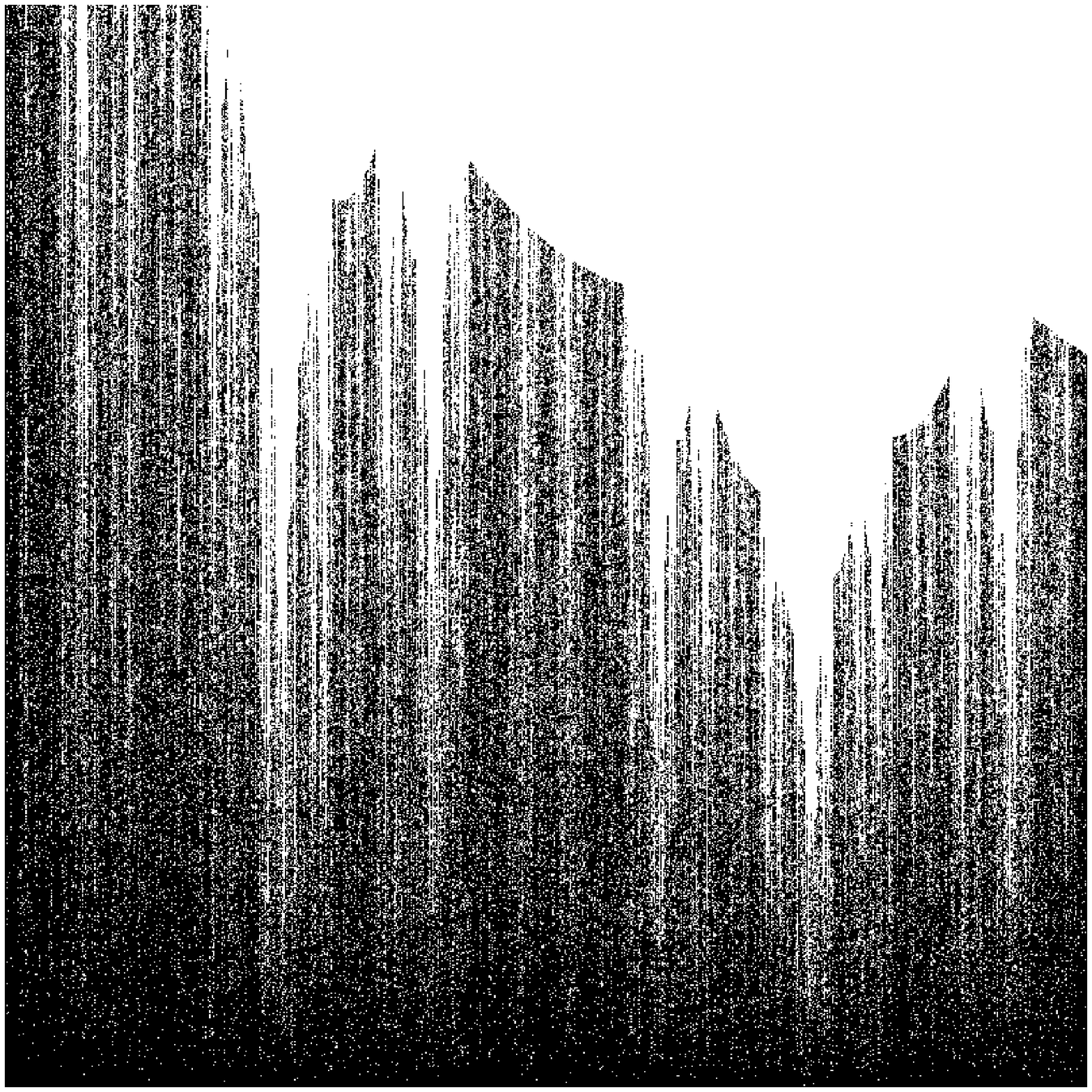,width=6cm}}

(b)

\end{center}
\caption{\label{fignnriddle}
The white regions show the local basin of attraction of
(a) a chaotic attractor in $N$ and (b) a hyperbolic periodic attractor in $N$.
In case (a) the basin is riddled whereas in (b) the basin is open.
Even though
the two sets look very similar; they differ fundamentally on a finer scale.
Points in black are attracted
to $N$ without leaving $y>0.1$, while those in white cross
the threshold $y>0.1$ before going
to $N$. The coordinate region depicted by the figures is given
by $x=[0,1]$, $y=[0,0.1]$,
and the parameters are fixed at
$(r,s,\nu,a,b)=(3.886,-0.3,0.545,-1,1)$
and $(r,s,\nu,a,b)=(3.88605,-0.3,0.545,-1,1)$
in (a) and (b) respectively.}
\end{figure}

This transition to riddled basins will in general be ``blurred''
as with the blowout bifurcation. Figure \ref{fignnriddle} shows
an example of basins of attraction of attractors within $N$ computed
for the map (\ref{eqmap}) at two different but close values of $r$. In
one case the attractor in $N$ is periodic and in the other a chaotic
attractor in $N$ with a riddled
basin. Between these two values there will be a non-normal
riddling bifurcation (or more precisely, a set of such bifurcations).
We note that despite their apparent similarity, the sets in
Figs.\ \ref{fignnriddle}(a) and
\ref{fignnriddle}(b) are fundamentally
different. To see this we show details in
Figs.\ \ref{figamplified}(a) and  \ref{figamplified}(b),
where the hyperbolic periodic attractor in $N$ clearly shows the presence of
open sets.

\begin{figure}
\begin{center}

(a)

\hspace{5mm}


(b)

\end{center}
\caption{\label{figamplified}
Detail of Figure~{\protect \ref{fignnriddle}}(a) and (b).
Panel (b) clearly shows the presence of open sets (the existence of the open
set can
also be inferred by noting that there is a linearly stable periodic point
at the centre of the bottom of this figure).  The coordinate regions
depicted here is given by
$x=[0.49,0.51]$, $y=[0,0.05]$. }
\end{figure}

\section{Non-normal parameters: theory}
\label{sectheory}

We explain the new observed effects of non-normal blowout, in-out
intermittency and non-normal riddling in the context of a recent
conjecture of Barreto {\em et al.}\, \cite{Bar&al97} on the parameter
dependence of chaotic attractors.

\subsection{Fragile chaos and the windows conjecture}

Barreto {\em et al.}\, consider a class of chaotic attractors that are
{\em fragile},
i.e.\ attractors that are not structurally stable under perturbations, but that
nonetheless persist for a large measure of nearby parameter values. Such
attractors are thought to be present in a large number of physically
important systems.
More precisely, Barreto {\em et al.}\, suppose that a mapping $h$ has a
chaotic attractor $A$ possessing $k\geq 0$ positive LEs.
Suppose that $h'$ is a mapping that is $C^1$ close to $h$ and such that
almost all points in a neighbourhood of $A$ are attracted to periodic
orbits. Then they say $A$ is {\em dispelled} for $h'$. They define
the attractor $A$ as being {\em fragile} if $A$ is dispelled for
$C^1$-arbitrarily close $h'$.

In particular, their main conjecture (the {\em windows conjecture}) for
one parameter and one positive LE, suggests that there will be a dense
set of nearby parameter values at which the attractors are periodic.
This set of parameter values is called the {\em windows set}.

For the model (\ref{eqmap}) that we consider, there are two deep
results about the logistic map that mean that the chaotic attractors
of the map on $N$ are really fragile in the above sense. Namely:

\begin{theorem}(see \cite{Jak81,Gra&Swi97})\label{thmlogistic}
Consider the mapping $f_a(x)=ax(1-x)$ from $[0,1]$ to itself. There is
a positive measure subset $I_c$ of $[0,1]$ such that $a\in I_c$ implies
that $f_a$ has a chaotic attractor with absolutely continuous invariant measure
and one positive LE. The complement of this set contains an open and dense
subset $I_p\subset[0,1]$ such that $a\in I_p$ implies that almost all
points are attracted to a stable periodic orbit.
\end{theorem}

\subsection{Attractors and invariant measures}

We briefly sketch some standard definitions and results that we need.
Suppose $f:\R^m\rightarrow\R^m$. If $A\subset \R^m$ is a compact invariant
set then we define
$$
\cW^s(A)=\{x\in\R^m~:~\omega(x)\subseteq A\}
$$
to be the {\em stable set} or {\em basin of attraction} of $A$; similarly
we define
$$
\cW^u(A)=\{x\in\R^m~:~\alpha(x)\subseteq A\}
$$
to be the {\em unstable set} of $A$.

Suppose that $\ell$ is Lebesgue measure on $\R^m$.
Following Milnor \cite{Mil85}, we say $A$ is an {\em attractor} if
$\ell(\cW)>0$. It is a {\em minimal attractor} if there is
no compact invariant proper subset $A'\subset A$ with
$\ell(\cW^s(A'))=\ell(\cW^s(A))$.

Suppose that $A$ is an attractor in the sense of Milnor. We say it
possesses a {\em natural measure} if there is an ergodic invariant measure
$\mu$ supported on $A$ such that almost all points in $\cW^s(A)$ are
generic for $\mu$; i.e.\ if almost all $x\in\cW^s(A)$ satisfy
$$
\lim_{n\rightarrow\infty}\frac{1}{n}\sum_{k=0}^{n-1}\phi(f^k(x))
=\int \phi\,d\mu.
$$
for all compactly supported $\phi\in C^0(\R^m,\R)$.

Note that if $A$ supports chaotic dynamics then typically there are
many other ergodic invariant measures supported on $A$; these are
the singular measures:
$$
\cM_A=\{\nu~\mbox{ ergodic invariant measures }~:~\supp(\nu)\subseteq A\}
$$
and include, for example, measures supported on
unstable periodic motions contained within $A$.

The basic result of Oseledec's theorem \cite{Ose68} is that LEs
exist and take a finite number of constant values
on a set of full $\nu$-measure for
any ergodic $\nu$. If $A$ has a natural measure as well as
singular measures we say the attractor is {\em chaotic} and the LEs to
almost all initial points are given by those for the natural measure;
however the convergence is non-uniform and shows arbitrarily long deviations
towards the LEs for any singular measure for $A$.
This causes riddling and several other
phenomena associated with chaotic behaviour in an invariant subspace.

As shown, for example, in \cite{Yam&Fuj84,Pik84,Ale&al92,AshBueSte96},
for a large class of attractors
one can classify an attractor in an invariant subspace according to
its transverse LEs according to where zero lies
relative to this spectrum. Suppose that $\mu_\nu$ is a family of
natural measures of attractors for $h_\nu$ and define
$$
\lambda_T(\nu)=\max\{\lambda_{\perp}(\mu_\nu))\}
$$
the most positive transverse LE for the natural measure at $\nu$.
Then $A_\nu=\supp(\mu_{\nu})$ is an attractor for $f_\nu$
if $\lambda_\nu<0$. It is
not an attractor if $\lambda_T(\nu)>0$. Parameter values $\nu_0$
where $\lambda_T$ takes both signs for $\nu$ arbitrarily close
to $\nu_0$ we refer to as {\em blowout points} of $A_\nu$
and these govern the bifurcation from the invariant subspace \cite{Ott&Som94}.

\subsection{Blowout with normal parameters}

The transverse LEs can be thought of as the characteristic exponents
for the linear skew product system
$$
\tilde{f}_\nu(x,v)=(h(x),M_\nu(x)v)
$$
where $M(x)$ is a normal derivative at $x$; i.e.\ such that
$$
D_{(x,0)}f_\nu \vect{u}{v}=\matr{*}{*}{0}{M_\nu(x)} \vect{u}{v},
$$
where $*$ denotes some matrix function of $x$. If we consider such a
system perturbed by normal parameters, the transverse LEs
are simply those of a perturbed cocycle and so in
particular they can vary continuously with any normal parameter (more
precisely, they will vary continuously on a generic set of perturbed
systems \cite{Arn&Con94}). If the LEs do vary
continuously then in particular $\lambda_T$ will vary continuously
and even
$$
\frac{d\lambda_T}{d\nu}\neq 0
$$
generically for $\lambda_T=0$; thus we can see that (at least
within an open set of smooth systems), generically the set of blowout
points has codimension one in parameter space. On one side of such a
normal blowout, $A_\nu$ is an attractor for $f_\nu$; on the other
side it is not.  When $A_\nu$ loses stability it may give rise to a
branch of on-off intermittent attractors (non-hysteretic or supercritical
scenario),
or there may be no nearby attractors (hysteretic or subcritical scenario), as
discussed by Ott \& Sommerer (although there are further
possibilities, as discussed in \cite{Ash&Ruc97}).  Note that the
hysteretic scenario may give rise to transient on-off intermittent
behaviour \cite{Xie&Hu96} near blowout.

\subsection{Blowout with non-normal parameters and the blowout set}

If we do not have normal parameters and have fragile chaos in the
invariant subspace, the attractors $A_\nu$ will typically collapse
over some dense set of windows and consequently $\lambda_\nu$ will
typically not vary continuously.

As $\lambda_T(\nu)$ does not vary continuously we cannot apply any
intermediate value theorem and so it is possible to pass from
$\lambda_T<0$ to $\lambda_T>0$ without passing through zero.
Therefore, we define
$$
I^+=\{\nu~:~\lambda_T(\nu)\geq 0\}, ~~I^-=\{\nu~:~\lambda_T(\nu)\leq 0\}
$$
and the {\em blowout set} to be
$$
I^0=\partial I^+\cap \partial I^-=\overline{I^+}\cap \overline{I^-}.
$$
For normal parameters, this is equivalent to the set of blowout
bifurcations.  The numerical evidence of Section~\ref{secmodel}
suggests that $I^0$ can be a fractal: more precisely we can use
Theorem~\ref{thmlogistic} to show the following.

\begin{corollary}
There is a map of the form (\ref{eqmap}) whose blowout set $I^0$ has
positive Lebesgue measure in parameter space.
\end{corollary}

\proof Consider any map that has linear form
$$
f_a(x,y) =(ax(1-x),(a-2ax)y)
$$
near the invariant subspace $y=0$. This is such that the tangential
and the transverse LEs are the same for all $a$.  On varying the
non-normal $a$, it is possible to see that $I^+=I_c$ and $I^-=I_p$ in
the notation of Theorem~\ref{thmlogistic}; the conclusions of this
theorem imply that $I^0=I^+$ and consequently the blowout set has
positive Lebesgue measure.

\eproof

In the more general case of $A$ being an attractor that is stuck on to $N$
with $A_0=A\cap N$
fragile then we conjecture that there are typically positive measure blowout
sets at loss of transverse stability. If $A_0$ is not fragile but instead
structurally stable (e.g. if it is uniformly hyperbolic) then the
blowout set has zero Lebesgue measure.
However, even in the case of a positive measure blowout set, the sparcity
of windows may mean that they are very difficult to observe even
in a region where they exist.

\section{In-out intermittency}
\label{secinout}

We now turn to a static (i.e.\ non-bifurcation) effect that is
associated with the blurred blowout. In particular, we characterise
a generalised form of on-off intermittency that we refer to
as in-out intermittency. This includes forms of on-off intermittency
where the attracting dynamics within the
invariant subspace does not need to be chaotic.

In the following, we will suppose we have a minimal Milnor attractor
$A$ for $f$  such that $A_0=A\cap N$ is non-empty and $A\cap N^c$
is also non-empty.

We write $\cA$ to be the set of invariant subsets of $A_0$ and
define
$$
\cA_0^{*}=\{B\subset A_0~:~ B\mbox{ is a minimal attractor for $f|_N$}\}
$$
and
$$
A_0^{*}=\overline{\cup_{B\in\cA^{*}} B}.
$$
This is a closed invariant subset of the invariant
set $A_0$.  We can think of $A_0^{*}$ as the set of points
that attract almost all points within $A^0$. If
\begin{equation}\label{eqonoff}
A_0^{*}=A_0
\end{equation}
then we say the attractor $A$ displays {\em on-off intermittency}. In
the more general case where
\begin{equation}\label{eqinout}
A_0^{*}\subseteq A_0
\end{equation}
we say that $A$ displays {\em in-out intermittency}, implying that
the attraction and repulsion to the invariant subspace are dominated
by different dynamics.
\\

\noindent {\bf Hypothesis} We will assume that if $f$ has a Milnor attractor
$A$ and $A_0=A\cap N$ is non-empty for some invariant manifold $N$
then $A_0$ is a Milnor attractor for $f|_N$.
\\

Note that if $A$ is an asymptotically stable attractor then this will
always hold; at present we have no measurable examples where this fails
to hold but are not aware of a sufficiently general setting where this
will always hold. Therefore we will leave it as a standing hypothesis.
Necessarily this means that $A_0^*$ will be non-empty, but it may be a
proper subset of $A_0$.

\subsection{Skew product systems}

The original examples of on-off intermittency and much theory
has been developed for skew product systems. Interestingly enough,
non-trivial in-out intermittency cannot occur in such a system,
as the following result implies.

\begin{lemma}\label{lemskewminimal}
Suppose that $f$ has skew product form (\ref{eqskewprod}) and $A$ is a
minimal Milnor attractor for $f$. Then $A_0=A\cap N$ is a minimal attractor
for $g=f|_{N}$.
\end{lemma}

\proof
Suppose $\Pi:\R^m\rightarrow N$ defined by $\Pi(x,y)=x$ is the
orthogonal projection onto $N$. If $f$ has form (\ref{eqskewprod})
then $\Pi\circ f=g\circ \Pi$ shows that $g$ is a factor of $f$. This
means that $A_0$ is compact and invariant and moreover,
$\cW^s(A_0)\cap N=\Pi(\cW^s(A))$; since the image of a set with
positive $m$ dimensional Lebesgue measure under $\Pi$ has
positive $n$ dimensional measure, this means that $A_0$ is
a Milnor attractor for $g$. Suppose that $A_0$ is not minimal for $g$.
Then there will be a positive measure subset of $\cW^s(A_0)\cap N$ that
converge to some proper compact subset $B\subset A_0$. Thus
there will be a positive measure subset of $\cW^s(A)$ that converge to
a set $B'$ with $\Pi(B')=B$ and $B'\subset A$.  Taking the closure of
$B'$, this contradicts the assumption that $A$ is minimal.

\eproof

The following theorem is a direct result of Lemma~\ref{lemskewminimal}.

\begin{theorem}
If $A$ is a minimal Milnor attractor for a skew product system and $A$ displays in-out
intermittency then it displays on-off intermittency.
\end{theorem}

\begin{figure}
\begin{center}
\mbox{
\psfig{file=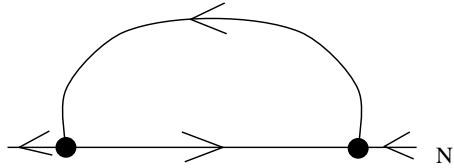,width=6cm}
}
\end{center}
\caption{\label{figheteroclinic}
Heteroclinic cycle consisting of two equilibria in $N$ connections
between them. If this forms part of an attractor then the only
minimal attractor within $N$ is the fixed point that is stable within
$N$.
}
\end{figure}

If $f$ is not a skew product then $A_0^*$ can be a proper subset of $A_0$.
For example, see the illustration in Figure~\ref{figheteroclinic} of
a heteroclinic network in $\R^2$ where the only minimal attractor in $N$ is
a fixed point.

The existence of in-out intermittency forces several consequences
in terms of LEs (where $\lambda_T(\mu)=\max \lambda_{\perp}(\mu)$
for an invariant measure $\mu$):

\begin{lemma}
Suppose that $A$ is an attractor displaying in-out intermittency
and $A_0^*$ has a natural measure $\mu^*$ for $f|_{N}$. Then
(i) $A_0$ is not uniquely ergodic,
(ii) $\lambda_T(\mu^*)\geq 0$ and
(iii) there exists a measure $\mu^\dag$ with support on
$A_0\setminus A_0^*$ such that $\lambda_T(\mu^\dag)\leq 0$.
\end{lemma}

\proof
(i) Is a trivial consequence of the fact that $A_0^* \neq A_0$.
(ii) Suppose that $\lambda_T(\mu^*)<0$; then $A_0^*$ is a
minimal attractor for $f$. (iii) Likewise, if this were not
the case then $A^0$ would be an attractor for $f$.

\eproof

As far as the dynamics on $A_0$ and the minimal attractors $A_0^*$ go
there seem to be many possibilities. In the numerical example, we see
examples  where $A_0$ is a chaotic repellor and $A_0^*$ is periodic.
There is no reason why $A_0^*$ should be composed of a single minimal
attractor, or indeed why $A_0^*$ itself should not be chaotic.

We now address the question of how to model the in-out intermittent
states themselves; we do this with the aid of a Markov model.

\subsection{A Markov model of in-out intermittency}

We motivate, describe and analyse a Markov chain model of in-out
intermittency that enables us to predict scalings associated with this
form of intermittency. In particular we consider the laminar
phases, corresponding to phases where the trajectory is below
a given threshold from the invariant submanifold.
Note that unlike the case for on-off
intermittency, we do not have a corresponding Fokker-Planck equation
which models this Markov model (cf \cite{Pik84}).

\begin{figure}
\begin{center}
\mbox{
\psfig{file=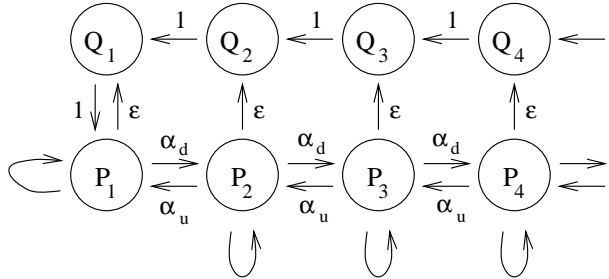,width=8cm}
}
\end{center}
\caption{\label{figchain}
Schematic diagram showing transitions in the Markov model for in-out
intermittency. The site $i$ can be thought of as indexing a distance
$\gamma^i$ from the invariant manifold for some $\gamma<1$ and so
we approach the invariant subspace on letting the index $i\rightarrow\infty$.
}
\end{figure}

Consider a Markov chain described by two semi-infinite chains of
states $P_i$ and $Q_i$, $i=0,1,2,\ldots$. We call the chain $P_i$
the ``in'' chain and $Q_i$ the ``out'' chain. We assign transition probabilities
as shown in Table~\ref{tabtransitions} and show the chain schematically in
Figure~\ref{figchain}. This model has three parameters
$$
\alpha_U, ~~ \alpha_D~~\mbox{ and }~~\epsilon
$$
which are all positive and such that $\alpha_U+\alpha_D+\epsilon\leq
1$. For simplicity we assume that $\alpha_U+\alpha_D+\epsilon=1$
and so the transition from $P_i$ to $P_i$ cannot happen. In this case
we can reparametrise using $\lambda\in\R$ by defining
$$
\alpha_U = \frac{1-\epsilon+\lambda}{2},~~
\alpha_D = \frac{1-\epsilon-\lambda}{2}.
$$
and note that $\lambda$ corresponds to a bias in the drift on
the ``in'' chain.

\begin{table}
$$
\begin{array}{c|cccccc}
 & \multicolumn{6}{c}{\mbox{To}}\\
         & P_{i-1} & P_i & P_{i+1} & Q_{i-1} & Q_{i} & Q_{i+1} \\
 \hline
 P_{i}&\alpha_U & 1-\alpha_U-\alpha_D-\epsilon & \alpha_D & 0 & \epsilon & 0\\
 Q_{i  } &   0     & 0   &  0    &  1   &  0   &   0     \\
\end{array}
$$
($i\geq 2$)
$$
\begin{array}{c|ccc}
 & \multicolumn{3}{c}{\mbox{To}}\\
         & P_{1}       & P_2      & Q_1 \\
 \hline
 P_{1}   &  1-\alpha_D -\epsilon & \alpha_D & \epsilon \\
 Q_{1}   &  1          &  0       & 0 \\
\end{array}
$$
\caption{\label{tabtransitions}
Probability of transition from the states $P_i$ and $Q_i$ for $i\geq 1$
in the Markov model of in-out intermittency. The $P_i$ model a biased
random walk that can leak onto a deterministic motion away from $\infty$.
At $i=1$ the $Q_i$ chain feeds into the $P_i$ chain.}
\end{table}

\paragraph{Equilibrium distributions}

We compute equilibrium probability distributions $p_i$ and $q_i$ of
the above model by solving the linear recurrence relations:
\begin{eqnarray*}
p_i & = & \alpha_U p_{i+1}+\alpha_D p_{i-1}+
(1-\alpha_U-\alpha_D-\epsilon)p_{i},~~i\geq 2\\
q_i & = & q_{i+1}+\epsilon p_i,~~i\geq 2\\
p_1 & = & \alpha_U p_{2}+q_1+(1-\epsilon-\alpha_D)p_1.
\end{eqnarray*}
Expressing in terms of the parameters $\lambda$ and $\epsilon$
we use an ansatz $p_i=\mu^i$ to obtain
\begin{eqnarray}
q_i & = & \frac{\epsilon}{1-\mu} \mu^i \\
p_i & = & \mu^i
\end{eqnarray}
where $\mu$ satisfies
$$
(1-\epsilon+\lambda)\mu^2-2\mu+(1-\epsilon-\lambda)=0.
$$
This quadratic equation has solutions
$$
\frac{1\pm\sqrt{\lambda^2 +2\epsilon- \epsilon^2}}{
1-\epsilon+\lambda}
$$
but we are only interested in normalised probability measures;
this means that the only physically relevant solution (with
$|\mu|\leq 1$) is
$$
\mu=\frac{1-\sqrt{\lambda^2 +2\epsilon- \epsilon^2}}{
1-\epsilon+\lambda}.
$$
For normalisation, the sum over all probabilities is unity
\begin{equation}
1=\sum_{i=1}^{\infty} C(p_i+q_i)=
C\frac{\epsilon+1-\mu}{(1-\mu)^2},
\end{equation}
giving
$C=\frac{(1-\mu)^2}{\epsilon+1-\mu}$. Now the asymptotic
ratio between time spent in the ``in'' and ``out'' phases,
$R_{io}$, which is a dynamical invariant, can be easily calculated as
\begin{equation}
R_{io}=
\frac{\sum_{i=1}^{\infty} C p_i}{\sum_{i=1}^{\infty} C q_i}=
\frac{1-\mu}{\epsilon}
\end{equation}
and the average value of the transverse variable $y$ can be similarly found
to be
\begin{equation}
<y>=\sum_{i=1}^{\infty} C (p_i+q_i) \gamma^{-i}=
\frac{(1-\mu)\gamma}{\gamma-\mu}.
\end{equation}

This quantity $R_{io}$ can be seen to be an invariant of the system up
to conjugacy and so is a quantity by which the intermittency
can be characterised. The case $R_{io}=\infty$
(equivalently $\epsilon=0$) corresponds to
when the trajectory stays on the ``in'' phase and hence
performs on-off intermittency.
In the case $R_{io}\ll 1$, the trajectory will spend long times
moving away from the invariant subspace on the ``out''
phase interspersed with short and necessarily fast-moving
contractions towards the invariant subspace.
We have seen examples of in-out intermittency
in the map (\ref{eqmap}) that show a wide range of
$0\leq R_{io}<\infty$.

Note that $\log \gamma$ corresponds to the
transverse LE $\lambda_T$ for the ``out''
phases. Usually for systems with normal parameters, the average laminar
phase as a function of $(r-r_c)$, where $r_c$ is the value of $r$
for which the blowout bifurcation occurs, scales as $<y>^{-1}$.
If there is a scaling between $<y>$ or $R_{io}$
and the normal parameters $\nu$, $a$, $b$ or $s$, then we can
go further and obtain $\mu$.

Therefore, we can use this model for any in-out intermittent attractor
by identifying
$$
\lambda_T,~~ \lambda,~\mbox{ and }~\epsilon,
$$
or equivalently, by measuring
$$
\lambda_T,~~R_{io},~~\mu~\mbox{ and }~<y>
$$
and computing $\gamma$, $\lambda$, $\mu$ and $\epsilon$ via
\begin{eqnarray*}
\mu&=&\frac{1-\sqrt{\lambda^2 +2\epsilon- \epsilon^2}}{1-\epsilon+\lambda}\\
R_{io}&=&\frac{\epsilon}{1-\mu}\\
<y>&=&\frac{(1-\mu)\gamma}{\gamma-\mu}\\
\lambda_T&=&\log\gamma.
\end{eqnarray*}

We can also calculate the average length of ``out'' phases, $L_{o}$ as
\begin{equation}
L_{o}=\sum_{i=1}^{\infty} C q_i=\frac{1}{1-\mu-\epsilon}.
\end{equation}

Note that it would be equally possible to have more than two chains
corresponding to, for example, more than one type of ``out'' dynamics.
Equally, the out dynamics could be chaotic and give rise to a biased
but two-directional random walk on the out-chain. There are
many possibilities and we do not examine all of these
in detail on this occasion, however we have observed such behaviours
in the model mapping (\ref{eqmap}).

\subsection{Scalings of laminar phases}

We now compute long time asymptotic properties of the scaling of laminar
phases for in-out intermittency by analysing the Markov model
proposed in the previous section.  We find that for small
$\epsilon$ this is very similar to that found for on-off intermittency;
namely for small times it decays according to a power law $n^{-3/2}$
whereas for larger times it decays much faster, namely exponentially.

To compute the distribution we consider a trajectory starting at
some site on the ``in'' chain and consider the distribution of
times as to when it returns for the first time to that site
in the  ``in'' chain.
Suppose we want to find the probability of the
trajectory returning for the first time after $2n$ steps.
This can occur either by the trajectory remaining in the ``in'' chain for
that excursion, or it can happen by `leaking' onto the $2n-2l$th
site further down the ``out'' chain after a time $2l$ and then propagating
up the ``out'' chain for the remaining time.

More precisely, let
$$
S_{2n}=P\{\mbox{first return is after time $2n+2$}\}
$$
where this is computed assuming that we start at a given site on the
``in'' chain and $P_{in}(2n+2)$ the probability of the first return
occur at time $2n+2$ always in the ``in'' chain.  Let
$$
T_{2n}(2l)=P\{\mbox{attains site $2l$ after a time $2n$, has not
returned or switched before then}\}.
$$
Then we can compute
$$
S_{2n}=P_{in} (T=2n+2)
+\epsilon\sum_{l=\lceil n/2\rceil}^{n-1}T_{2l}(2n-2l).
$$

Using a directed version of the probability of the first return and
the hitting time formulas \cite{Feller,Gri&Sti82}
given by
$$P_{in} (T=2n+2) =\frac{1}{n+1}\vect{2n}{n}(\alpha_D\alpha_U)^{n+1}$$
and
$$
T_{2n}(2l)=\frac{l}{n}\vect{2n}{n+l}(\alpha_D)^{n+l}(\alpha_U)^{n-l}
$$
respectively, we obtain, for $\alpha_U \ne 0$,
\begin{eqnarray*}
S_{2n} &=& \frac{1}{n+1}\vect{2n}{n}(\alpha_D\alpha_U)^{n+1}
+\epsilon\sum_{l=\lceil n/2\rceil}^{n-1} \frac{n-l}{l}\vect{2l}{n}
(\alpha_D)^{n}(\alpha_U)^{2l-n}\\
&=& \frac{1}{n+1}\vect{2n}{n}(\alpha_D\alpha_U)^{n+1}
+\epsilon\left(\frac{\alpha_D}{\alpha_U}\right)^n
\sum_{l=\lceil n/2\rceil}^{n-1} \frac{n-l}{l}\vect{2l}{n} (\alpha_U)^{2l}.
\end{eqnarray*}
It now remains to find the asymptotic behaviour of this expression.
Using Stirling's formula $n!\sim n^{n+\frac{1}{2}}e^{-n}\sqrt{2\pi}$
we have
$$
\vect{2n}{n}\sim \frac{4^n}{\sqrt{n\pi}}
$$
which upon taking $n\rightarrow\infty$ gives
\begin{eqnarray*}
S_{2n} &\sim &
\alpha_U\alpha_D\, n^{-\frac{3}{2}}\frac{(4\alpha_U\alpha_D)^n}{\sqrt{\pi}}\\
&& +\epsilon
\left(\frac{\alpha_D}{\alpha_U}\right)^n \frac{1}{n^{n}\sqrt{2\pi n}}
\sum_{l=\lceil n/2\rceil}^{n-1} \frac{n-l}{l} (2l)^{2l+\frac{1}{2}}
(2l-n)^{-2l+n-\frac{1}{2}} (\alpha_U)^{2l}\\
&=& I_1+I_2.
\end{eqnarray*}
We can approximate $I_2$ by
\begin{eqnarray*}
I_2 &\sim & \epsilon
\left(\frac{\alpha_D}{\alpha_U}\right)^n \frac{1}{n^{n}\sqrt{2\pi n}}
\int_{x=\lceil n/2\rceil}^{n} \frac{n-x}{x} (2x)^{2x+\frac{1}{2}}
(2x-n)^{-2x+n-\frac{1}{2}} (\alpha_U)^{2x}\,dx\\
\end{eqnarray*}
which by changing the integration
variable to $x=n(1+y)/2$ becomes
\begin{eqnarray*}
I_2 &\sim& \epsilon\left(\frac{\alpha_D}{\alpha_U}\right)^n
\frac{\sqrt{n}}{\sqrt{2\pi}}\int_{0}^{1} \frac{1-y}{\sqrt{(1+y)y}}
(1+y)^{n(1+y)} y^{-ny} (\alpha_U^n)^{1+y}\,dy\\
&\sim& \epsilon\left(\frac{\alpha_D}{\alpha_U}\right)^n\frac{\sqrt{n}}{\sqrt{2\pi}}
\int_0^1 q(y) e^{np(y)}\,dy
\end{eqnarray*}
where $p(y)=(1+y)\log((1+y)\alpha_U)-y\log y$
and $q(y)=\frac{1-y}{\sqrt{y(1+y)}}$.
Note that $p(y)$ has a single maximum
at $y_m\in[0,1]$ given by solving
$$
p'=\log(\alpha_U(1+y))-\log y =0
$$
and $p''(y)=\frac{1}{1+y}-\frac{1}{y}$.
We need at this point to distinguish between two cases for $\alpha_U$.

\paragraph{Case of $0<\alpha_U<\frac{1}{2}$}
In this case, $y_m=\frac{\alpha_U}{1-\alpha_U}$
and the maximum $y_m$ is in the interior of the range of
integration. Laplaces' method gives
$$
\int_0^1 q(y)e^{np(y)}\,dy\sim q(y_m)e^{np(y_m)}
\frac{1}{\sqrt{-2\pi np''(y_m)}}
$$
asymptotically as $n\rightarrow\infty$ and so we can estimate
$$
I_2\sim\frac{\epsilon}{2\pi}\frac{2\alpha_U-1}{\alpha_U-1}
\left(\frac{\alpha_D}{1-\alpha_U}\right)^{n}
$$
to leading order. Note that this implies that $I_1\ll I_2$
and $I_2$ gives exponential fall off of $S_{2n}$ in $n$ for large $n$.
Note however the factor $\epsilon$ in $I_2$ means that $I_1$ may dominate
the statistics up to a moderately large values of $n$.

\paragraph{Case of $1>\alpha_U>\frac{1}{2}$}
In this case $y_m=1$ and $p'(1)>0$. This means that the leading term in
$I_2$ is given by
$$
I_2\sim \epsilon n^{-\frac{3}{2}}(4\alpha_D\alpha_U)^n
$$
as $n\rightarrow\infty$. This is the same scaling with $n$ as
for $I_1$ and so it will appear to scale exactly as for on-off intermittency.
Thus in this case
$$
S_{2n}\sim n^{-\frac{3}{2}}(4\alpha_U\alpha_D)^n
$$
in the limit of large $n$. Rewriting this as
$$
S_{2n}\sim n^{-\frac{3}{2}}\exp(n \log(4\alpha_U\alpha_D))
$$
we note that for any $\epsilon>0$
$$
0<4\alpha_U\alpha_D=(1-\epsilon)^2-\lambda^2<1
$$
and so $S_{2n}$ scales as $n^{-3/2}$ up to the point
where exponential decay sets in.

\paragraph{Other cases} In the case $\alpha_U=\frac{1}{2}$ we note that
$y_m=1$ and $p'(1)=0$; we do not consider this case in detail.
In the case $\alpha_U=0$ we have a deterministic propagation down the ``in''
chain and it is easy to compute that
$$
S_{2n}\sim \epsilon (1-\epsilon)^n
$$
i.e.\ we have exponential decay only, and no $n^{-\frac{3}{2}}$
scaling. This case is the limiting case where $\lambda\rightarrow \epsilon-1$.

\begin{figure}
\begin{center}
\mbox{\psfig{file=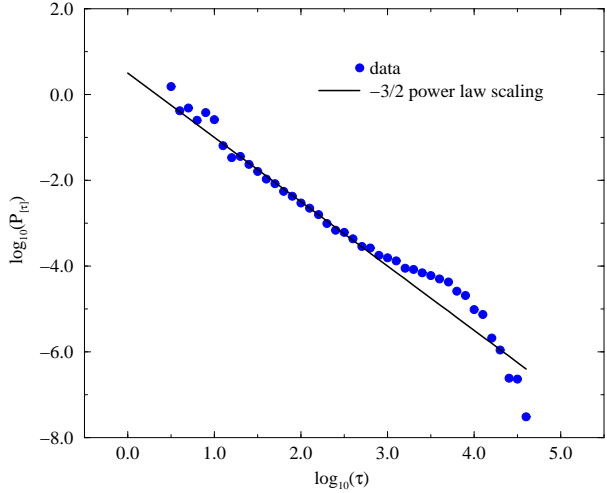,width=8cm}}
\end{center}
\caption{\label{laminar}
Scaling law for proportion $P$ of laminar phases of length $\tau$ for
in-out intermittency at $r=3.8800045$
and $\nu=1.82$, $a=-1$, $s=-0.3$ and $b=-1$. This case corresponds
to a case for which $\alpha_U<\frac{1}{2}$. Note the $n^{-3/2}$ fall off
changing to exponential for larger times.}
\end{figure}

\subsection{Distribution of laminar phases: summary}

We note that if $\epsilon>0$ for any value of $\lambda$
the distribution $S_{2n}$ of laminar phases of length $2n$ always
decays exponentially with $n$. For large $n$, we consider
two cases in detail. In the case where $\alpha_U>\frac{1}{2}$ we have
$$
S_{2n}\sim n^{-3/2}e^{n\log(4\alpha_U\alpha_D)}
$$
($\log(4\alpha_U\alpha_D)=\log((1-\epsilon)^2-\lambda^2)\leq 0$
with equality if and only if $\lambda=\epsilon=0$).

In the case of $\alpha_U<\frac{1}{2}$, or equivalently when
$\epsilon>\lambda$, there is an asymptotic scaling of the form
$$
S_{2n}\sim \frac{\alpha_U\alpha_D}{\sqrt{\pi}} 
n^{-3/2}e^{n\log(4\alpha_U\alpha_D)}+
\frac{\epsilon}{2\pi}\frac{2\alpha_U-1}{\alpha_U-1}\ e^{n\zeta}
$$
where $\log(4\alpha_U\alpha_D)<\zeta<0$. This causes the exponential
tail to continue to much higher $n$, and indicates the presence of
a higher proportion of very long laminar phases. The shoulder
at high $n$ in Fig.\ \ref{laminar} is caused by the contribution
of $I_2$ and is a clear indicator of in-out intermittency. Pure on-off
intermittency, in the noise-free case, has a scaling law which
is always convex in the log-log plot.
It is also possible to use these results to
understand how the scaling of laminar phases
varies on changing LEs in the intermittency.

\section{Other bifurcations on varying a non-normal parameter}
\label{secbifs}

The bifurcation to riddling (and associated bifurcation to bubbling)
has been discussed in detail by
\cite{Ott&al93,Ott&al94,Lai&al96,Ven&al96,Ven&al96b},
in the case where a normal parameter is varied.

In this section we can see how basin riddling can appear on varying a
non-normal parameter; this occurs via a breakdown of fragile chaos of
an attractor within the invariant subspace.

\subsection{Bifurcation to riddling}

If we vary a non-normal parameter we expect riddled basins
to typically appear and disappear in the following way;

Suppose there is a $\nu_0$ such that
\begin{itemize}
\item[(a)] There is a branch of attracting periodic orbits
(or more generally, uniquely ergodic attractors $B_\nu\subset N$
for $\nu>\nu_0$).
\item[(b)] There is a large measure subset of $\nu<\nu_0$
where there exist a branch of attractors $C_\nu\subset N$ such that
$B_\nu$ are in a neighbourhood of the $C_\nu$ and the basins of
the $C_\nu$ are locally riddled.
\end{itemize}
Then we say there is a {\em non-normal bifurcation to riddling}
at $\nu_0$. Note that (b) holds if there is a transversely unstable
measure supported on $C_\nu$, subject to certain regularity
conditions. Some of these transversely unstable measures will
continue through the bifurcation to give rise to complicated
repelling sets for $\nu>\nu_0$.

We expect this new bifurcation is in fact a more typical way
to create riddled basins on varying a non-normal parameter than
the normal parameter scenario discussed in \cite{Ott&al94}.
An example of the basins of attraction before and after such
a bifurcation is shown in Figure \ref{fignnriddle}.

\subsection{Other transitions.}

There are many other bifurcations that have been studied in the case
of normal parameters and which will carry across to non-normal parameter
systems, modulo the fact that if the chaotic invariant set in $N$ is fragile
then these transitions will happen on sets that may be considerably large than
codimension one.

For example, the transition to bubbling \cite{Ven&al96,Ven&al96b}, an attractor
becoming unstuck \cite{Ash95} and a transition to cycling chaos \cite{Ash&Ruc97}
may all appear in a non-normal setting.

Moreover, the case $\epsilon\rightarrow 0$ in the Markov model for in-out
intermittency will correspond to a transition from on-off to in-out
intermittency.

\section{Discussion}
\label{secdiscuss}

The motivation for this paper came from previous work of the authors
\cite{Cov&al97} that investigated a truncated PDE model of an
axisymmetric mean field dynamo model.
Bifurcation and dynamical effects came to light that did not fit into the
usual setting of a blowout bifurcation; this led us to consider these
effects in the simple mapping (\ref{eqmap}).

In this paper we have addressed two main questions; firstly, what happens to
a blowout bifurcation in non-skew product systems
on varying the dynamics by a non--normal parameter
within the invariant subspace
and secondly, how to characterise the
dynamics of what we call ``in-out'' intermittency which we suggest
should be a commonly observable type of dynamics in realistic models with
invariant subspaces.
We have also raised a number of other
questions relating to riddling and other bifurcations and discussed these
in passing.

We have tried to give a
general characterisation of in-out intermittency, have contrasted it
to on--off intermittency and have related its likelihood
to the non--skew product nature of the system and
the non--normal nature of its parameters.  We have
proposed that this type of behaviour
can be well modelled by a simple Markov chain and
have used this to obtain quantitative measures of this state (notably
the ratio $R_{io}$ of ``in''- to ``out''- states, which is a dynamical
invariant).

An important signature of in--out intermittency is the presence of
intervals with exponential growth in the ``out'' variables
with a constant rate,
over many orders of magnitude.
While this occurs, the ``in'' variables closely shadow the
periodic attractor in the invariant submanifold,
thus giving a clear distinction
from on--off intermittency where the on and off phases are
not so clearly differentiated.

We shall show in a future publication that in--out intermittency
can account for the behaviour observed in the study
of mean field dynamo models (modelled on partial differential
equations) and referred to as ``icicle'' intermittency
by Brooke \cite{Bro97}, who uses a skew product model to investigate this
(see also Tworkowski {\em et al.}\, \cite{twor} and
Brooke {\em et al.}\, \cite{Betal}) as well as in the truncated versions
of related models by the authors \cite{Cov&al97}.
Similarly this type of behaviour is
related to the ``new type of intermittency''
reported by Hasegawa {\em et al.}\, \cite{Has&al97}
in a ring of phase-locked loops.

There are still few rigorous results on what has been called
``fragile chaos'' \cite{Bar&al97}.  Further progress in the rigorous
classification of non-normal blowouts is clearly hampered by the lack
of these results in what is a very difficult area of analysis.

\section*{Acknowledgements}

PA is partially supported by a Nuffield ``Newly appointed
science lecturer'' grant and EPSRC grant
K77365.
EC is supported by grant BD/5708/95 -- Program PRAXIS
XXI, from JNICT -- Portugal.
RT benefited from PPARC UK Grant No. H09454. This research
also benefited from the EC Human Capital and Mobility (Networks) grant
``Late type stars: activity, magnetism, turbulence'' No. ERBCHRXCT940483.




\section*{References}
\begin{thebibliography}{99}

\bibitem{Ale&al92}
Alexander, J., Kan, I., Yorke, J.\ and You., Z.,
\newblock Riddled Basins.
\newblock {\em Int.\ Journal of Bifurcations and Chaos} {\bf 2}:795--813 (1992).

\bibitem{Arn&Con94}
Arnold, L. and Cong, N.\ D.,
\newblock Generic properties of Lyapunov exponents.
\newblock {\em Random and Computational Dynamics} {\bf 2}:335--345 (1994).

\bibitem{Ash95} Ashwin, P.,
\newblock Attractors stuck on to invariant subspaces.
\newblock {\em Phys. Lett. A} {\bf 209}:338--344 (1995).

\bibitem{Ash97} Ashwin, P.,
\newblock Cycles homoclinic to chaotic sets; robustness and resonance.
\newblock {\em Chaos} {\bf 7}:207--220 (1997).

\bibitem{AshAstNic97} Ashwin, P., Aston, P.\ and Nicol, M.,
\newblock On the unfolding of a blowout bifurcation.
\newblock {\em Physica D} {\bf 111}:81--95 (1997).

\bibitem{AshBueSte94}Ashwin, P., Buescu, J.\ and Stewart, I.,
\newblock Bubbling of attractors and synchronisation of oscillators.
\newblock {\em Phys.\ Lett.\ A} {\bf 193}:126--139 (1994).

\bibitem{AshBueSte96} Ashwin, P., Buescu, J.\ and Stewart, I.,
\newblock From attractor to chaotic saddle: a tale of transverse
instability.
\newblock {\em Nonlinearity} {\bf 9}:703--737 (1996).

\bibitem{Ash&Ruc97}
Ashwin, P. and Rucklidge, A.\ M.,
\newblock Cycling chaos: its creation, persistence and loss of stability
in a model of nonlinear magnetoconvection.
\newblock Technical Report 97/26, Dept. of Maths and Stats, University
of Surrey (1997).

\bibitem{Bar&al97}
Barreto, E., Hunt, B., Grebogi, C.\ and Yorke, J.,
\newblock From high dimensional chaos to stable periodic orbits:
the structure of parameter space.
\newblock {\em Phys.\ Rev.\ Lett.} {\bf 78}:4561--4 (1997).

\bibitem{Bro97}
Brooke, J. M.,
\newblock Breaking of equatorial symmetry in a rotating system: a spiralling
intermittency mechanism.,
\newblock {\em Europhysics Letters} {\bf 37}:171--176 (1997).

\bibitem{Betal} Brooke, J.M., Pelt, J., Tavakol, R.\ and Tworkowski, A.,
\newblock Grand minima and equatorial symmetry
breaking in axisymmetric dynamo models.
\newblock {\em A\&A}, in press (1998).

\bibitem{Cov&al97}
Covas, E., Ashwin, P.\ and Tavakol, R.,
\newblock Non-normal parameter blowout bifurcation: an example in a
truncated mean field dynamo model.
\newblock {\em Phys. Rev. E}, {\bf 56}:6451--8 (1997).

\bibitem{Gra&Swi97}
Grazcyk, J.\ and Swiatek, G.,
\newblock Hyperbolicity in the real quadratic family.
{\em Annals of Math}, {\bf 54}:1--52, (1997).

\bibitem{Gri&Sti82}
Grimmett, G.\ and Stirzaker, D.,
\newblock {\em Probability and random processes.}
\newblock Oxford University Press, 1982.

\bibitem{Feller}
Feller, W.,
\newblock {\em An Introduction to Probability Theory and Its
Applications, Vol.\ 1, 3$^{rd}$ ed.}
\newblock John Wiley \& Sons, 1968.

\bibitem{Has&al97}
Hasegawa, A., Komuro, M.\ and Endo T.,
\newblock {\em A new type of intermittency from a ring of four
coupled phase-locked loops.}
\newblock Proceedings of ECCTD'97, Budapest, Sept. 1997
(sponsored by the European Circuit Society).

\bibitem{Hun&Ott96}
Hunt, B.\ and Ott, E.,
\newblock Optimal periodic orbits of chaotic systems.
\newblock {\em Phys.\ Rev.\ Lett.} {\bf 76}:2254--57 (1996).

\bibitem{Jak81}
Jakobson, M. V.,
\newblock Absolutely continuous invariant measures for one-parameter
families of one-dimensional maps.
\newblock {\em Comm.\ Math.\ Phys.} {\bf 81}:39--88 (1981)

\bibitem{Kat&Has95} Katok, A.\ and Hasselblatt, B.,
\newblock {\em Introduction to the Modern Theory of Dynamical Systems}
\newblock (Encyclopedia of Mathematics and its Applications {\bf 54}).
Cambridge University Press, 1995.

\bibitem{Lai&Gre96} Lai, Y-C.\ and Grebogi, C.,
\newblock Characterising riddled fractal sets.
\newblock {\em Phys.\ Rev.\ E.} {\bf 53}:1371-58 reference 3 (1996).

\bibitem{Lai&al96} Lai, Y-C., Grebogi, C., Yorke, J.\ A.\
and Venkataramani, S.\ C.,
\newblock Riddling bifurcation in chaotic dynamical systems.
\newblock {\em Phys.\ Rev.\ Lett.} {\bf 77}:55-58 (1996).

\bibitem{Mil85}
Milnor, J.,
\newblock On the concept of attractor.
\newblock {\em Commun. Math. Phys.} {\bf 99}:177--195 (1985);
\newblock Comments {\em Commun. Math. Phys.} {\bf 102}:517--519 (1985).

\bibitem{Ose68}
Oseledec, V. I.,
\newblock A multiplicative ergodic theorem: Lyapunov characteristic
numbers for dynamical systems.
\newblock {\em Trans. Mosc. Math. Soc.} {\bf 19}:197--231 (1968).

\bibitem{Ott&al93}
Ott, E., Sommerer, J.\ C., Alexander, J., Kan, I. and Yorke, J.\ A.,
\newblock Scaling behaviour of chaotic systems with riddled basins.
\newblock {\em Phys. Rev. Lett.} {\bf 71}:4134--4137 (1993).

\bibitem{Ott&al94}
Ott, E., Sommerer, J.\ C., Alexander, J., Kan, I. and Yorke, J.\ A.,
\newblock A transition to chaotic attractors with riddled basins.
\newblock {\em Physica D} {\bf 76}:384--410  (1994).

\bibitem{Ott&Som94}
Ott, E.\ and Sommerer, J.,
\newblock Blowout bifurcations: the occurrence of riddled basins and on-off
intermittency.
\newblock {\em Phys.\ Lett.\ A} {\bf 188}:39--47 (1994).

\bibitem{Pik84}
Pikovsky, A.\ S.,
\newblock On the interaction of strange attractors.
\newblock {\em Z. Phys. B}, {\bf 55}:149--154, (1984).

\bibitem{Pla&al93}
Platt, M., Spiegel, E.\ and Tresser, C.,
\newblock On-off intermittency; a mechanism for bursting.
\newblock {\em Phys.\ Rev.\ Lett.} {\bf 70}:279--282 (1993).

\bibitem{Tav&Ell88}
Tavakol, R.\ K.\ and Ellis G.\ F.\ R.,
\newblock On the question of cosmological modelling.
\newblock {\em Phys.\ Lett.\ A} {\bf 130}:217--223 (1988).

\bibitem{twor} Tworkowski, A., Tavakol, R., Brandenberg, A., Moss, D. and
Tuominnen, I.,
\newblock Intermittent behaviour in axisymmetric mean field
dynamo models.
\newblock {\em Mon. Not. R. Astro Soc.}, in press (1998).

\bibitem{Yam&Fuj84}
Yamada, T.\ and Fujisaka, H.,
\newblock Stability theory of synchronised motion in coupled-oscillator
systems.
\newblock {\em Prog. Theor. Phys.} {\bf 70}:1240-8 (1984).

\bibitem{Ven&al96}
Venkataramani, S.\ C., Hunt, B.\ R., Ott, E., Gauthier, D.\ J.\ and
Bienfang, J.\ C.,
\newblock Transitions to bubbling of chaotic systems.
\newblock {\em Phys.\ Rev.\ Lett.} {\bf 77}:5361--4 (1996).

\bibitem{Ven&al96b}
Venkataramani, S.\ C., Hunt, B.\ R.\ and Ott, E.,
\newblock Bubbling transition.
\newblock {\em Phys Rev E} {\bf 54}:1346-60 (1996).

\bibitem{Xie&Hu96}
Xie F. and Hu, G.,
\newblock Transient on-off intermittency in a coupled map lattice system.
\newblock {\em Phys.\ Rev.\ E.} {\bf 53}: 1232--35 (1996).

\end{thebibliography}
\end{document}